\documentstyle[epsfig,natbib2,natbibmnfix]{mn}

\newcommand{\be}{\begin{equation}}
\newcommand{\ee}{\end{equation}}

\def\ltsima{$\; \buildrel < \over \sim \;$}
\def\simlt{\lower.5ex\hbox{\ltsima}}
\def\gtsima{$\; \buildrel > \over \sim \;$}
\def\simgt{\lower.5ex\hbox{\gtsima}}

\title[Transformations of CDM Halos]{Galaxy-Induced Transformation
  of Dark Matter Halos}

\author[Abadi et al ]{Mario G. Abadi $^{1}$, Julio F. Navarro$^{2}$,
Mark Fardal$^{3}$, Arif Babul$^{2}$ and Matthias Steinmetz$^{4}$
\\
$^{1}$ Observatorio Astron\'omico de C\'ordobai and CONICET, C\'ordoba, Argentina\\
$^{2}$Department of Physics and Astronomy, University of Victoria, Victoria, BC V8P 5C2, Canada\\
$^{3}$Astronomy Department, University of Massachusetts at Amherst, MA 01003, US\\
$^{4}$Astrophysikalisches Institut Potsdam, An der Sternwarte 16, Potsdam 14482, Germany\\
}

\begin{document}

\date{}
\pubyear{2008}
\maketitle
\label{firstpage}

\begin{abstract}
We use N-body/gasdynamical cosmological simulations to examine the
effect of the assembly of a central galaxy on the shape and mass
profile of its surrounding dark matter halo. Two series of simulations
are compared; one that follows only the evolution of the dark matter
component of individual halos in the proper $\Lambda$CDM cosmological
context, and a second series where a baryonic component is added and
followed hydrodynamically. The gasdynamical simulations include
radiative cooling but neglect the formation of stars and their
feedback. The efficient, unimpeded cooling that results leads most
baryons to collect at the halo center in a centrifugally-supported
disk which, due to angular momentum losses, is too small and too
massive when compared with typical spiral galaxies. This admittedly
unrealistic model allows us, nevertheless, to gauge the maximum effect
that galaxies may have in transforming their surrounding dark halos.
We find, in agreement with earlier work, that the shape of the halo
becomes more axisymmetric: post galaxy assembly, halos are transformed
from triaxial into essentially oblate systems, with well-aligned
isopotential contours of roughly constant flattening ($\langle
c/a\rangle \sim 0.85$).
Halos always contract as a result of galaxy assembly, but the effect
is substantially less pronounced than predicted by the traditional
``adiabatic contraction'' hypothesis. The reduced contraction helps to
reconcile $\Lambda$CDM halos with constraints on the dark matter
content inside the solar circle and should alleviate the long-standing
difficulty of matching simultaneously the scaling properties of galaxy
disks and the galaxy luminosity function. The halo contraction we
report is also less pronounced than found in earlier simulations, a
disagreement that suggests that halo contraction is not solely a
function of the initial and final distribution of baryons. Not only
{\it how much} baryonic mass has been deposited at the center of a
halo matters, but also the {\it mode} of its deposition. It might
prove impossible to predict the halo response to galaxy formation
without a detailed understanding of a galaxy's detailed assembly
history.
\end{abstract}

\section{Introduction}
\label{sec:intro}

Over the past couple of decades, cosmological N-body simulations have
established a number of important results regarding the structure of
cold dark matter (CDM) halos. In particular, broad consensus holds
regarding the overall shape of the halo mass distribution and its
radial mass profile. CDM halos are distinctly triaxial systems
supported by anisotropic velocity dispersion tensors
\citep{Frenk1988,Jing2002,Allgood2006,Hayashi2007,Bett2007}, a result
that reflects the highly aspherical nature of
gravitationally-amplified density fluctuations as they become non
linear and assemble into individual halos. Despite their chaotic
assembly, the spherically-averaged mass profile of CDM halos is nearly
``universal'', and can be approximated fairly accurately, regardless
of halo mass or redshift, by scaling a simple formula
\citep[see, e.g.,][]{Navarro1996,Navarro1997,Navarro2004,Navarro2008}.

One major limitation is that these results have been obtained from
simulations that neglect the presence of the baryonic, luminous
component of galaxy systems. Although this may be a suitable
approximation in extremely dark matter-dominated systems, it is
expected to fail in regions where baryons contribute a substantial
fraction of the mass, as is the case in the luminous regions of normal
galaxies like the Milky Way. The response of the dark halo to the
assembly of a galaxy is thus a crucial ingredient of models that
attempt to interpret observational constraints in terms of the
prevailing CDM paradigm.

There has been extensive work, both numerical and analytical, on the
modification of the mass profile of a dark halo induced by the
assembly of a central galaxy.  Early results suggested that simple
analytical prescriptions based on the conservation of adiabatic
invariants gave an accurate description of the halo
response. Following the early work by \citet{BarnesandWhite1984},
\citet{Blumenthal1986} devised a simple formula to link the dark mass
profiles before and after the assembly of a galaxy. Given the initial,
spherically-symmetric enclosed mass profiles of the dark matter,
$M_{\rm dm}^i(r)$, and baryons, $M_b^i(r)$, one may derive the final
dark mass profile, $M_{\rm dm}^f(r)$, once the final baryonic mass
profile, $M_b^f(r)$, is specified. The model assumes that dark matter
particles move on circular orbits before and after the contraction,
and that their initial, $r_i$, and final, $r_f$, radii are related by
the condition:
\begin{equation}
r_f \, [M_b^f(r_f)+M_{\rm dm}]=r_i \, [M_{\rm dm}+M_b^i(r_i)],
\label{eq:adcont}
\end{equation}
where $M_{\rm dm}=M_{\rm dm}(r_f)=M_{\rm dm}(r_i)$ is the dark mass
enclosed by each dark matter particle (i.e., no shell crossing).

Despite its simplicity, and the crude approximation on which it is
based, early N-body work
\citep[e.g.,][]{BarnesandWhite1984,Blumenthal1986,Jesseit2002} found
reasonable agreement between eq.~\ref{eq:adcont} and the results of
simulations, and helped to establish the ``adiabatic contraction''
formulation as the default procedure when considering the halo
response to the formation of a central galaxy.

Semianalytic models of galaxy formation have adopted this formulation
in order to link the (observed) dynamical properties of luminous
galaxies to the (predicted) properties of the dark halos that surround
them \citep[see,
  e.g.,][]{Mo1998,Cole2000,Dutton2005,Croton2006,somerville2008}.
This work has highlighted a number of potential problems when
attempting to reconcile the predicted properties of galaxies in the
$\Lambda$CDM cosmogony with observed scaling relations.

One particularly important challenge has been the inability of
semianalytic galaxy formation models to match simultaneously the zero
point of the Tully-Fisher (TF) relation and the galaxy luminosity
function. Successful models require that the rotation speed of disks
be of the order of the virial velocity of the halos they inhabit
\citep[see, e.g.,][]{Somerville1999,Croton2006}. However, models that
include adiabatic contraction typically predict disk rotation speeds
well in excess of the virial velocity.

\begin{figure*}
\begin{center}
\includegraphics[width=0.9\linewidth,clip]{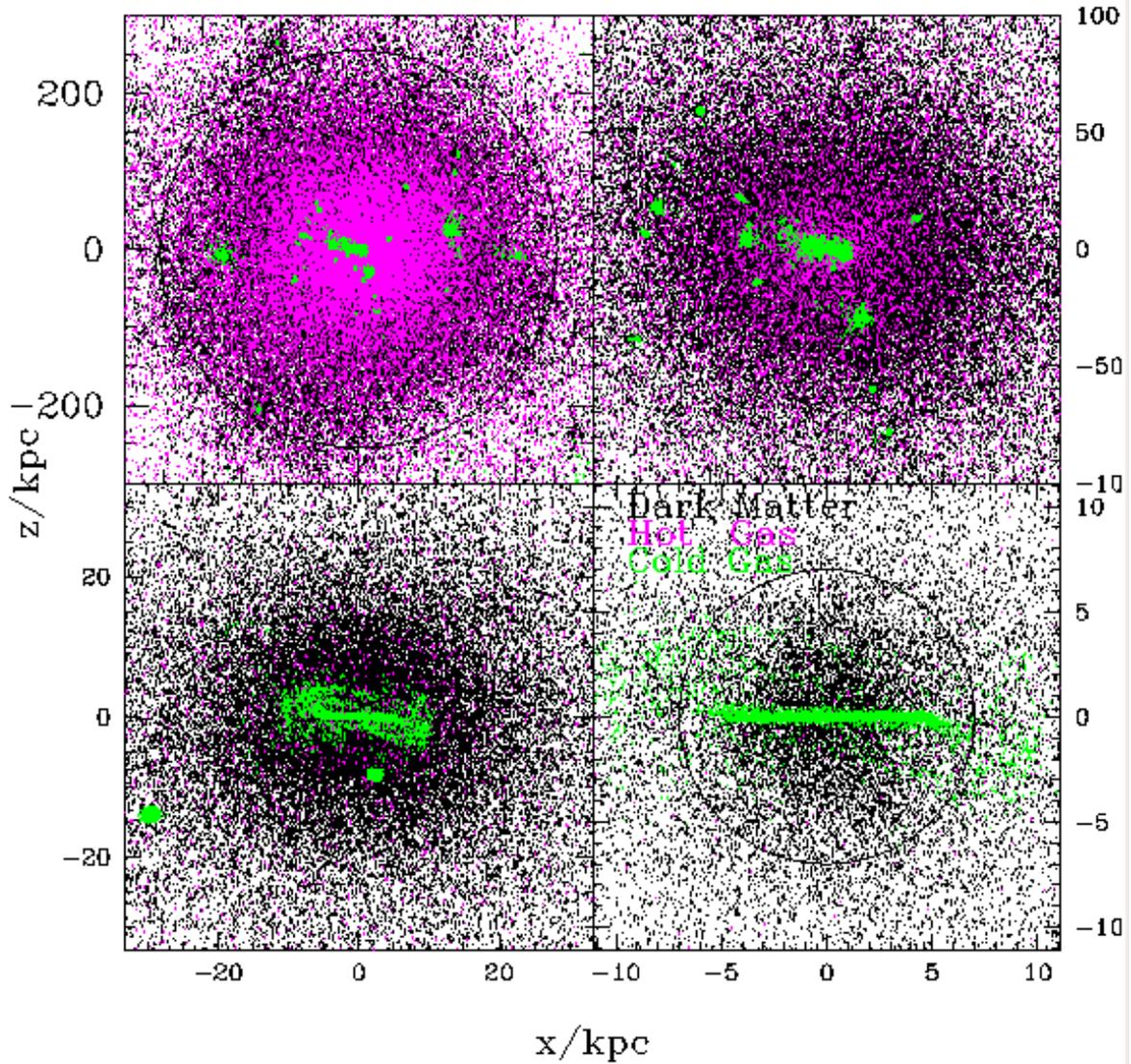}
\end{center}
\caption{Dark matter (black) and gas (colored) particles for halo
  S02h. The color of a gas particle reflects its temperature,
  $T<10^{4.5}$K (green) or $T>10^{4.5}$K (magenta). Each panel zooms
  into the center of the system by consecutive factors of 3 in radius
  (see axis labels). The circle in the top-left panel shows the virial
  radius. The circle in the bottom-right panel shows the radius,
  $r_{\rm glx}$, used to define the central galaxy.  Note that the
  colder (green) gas component inhabits the center of the main halo
  and of its substructures, where it forms thin,
  centrifugally-supported disks. The hotter component (magenta) is
  distributed more or less uniformly across the main halo and makes up
  only $16\%$ of all the baryons within the virial radius.
\label{figs:xyall}}
\end{figure*} 

As discussed most recently by, for example,
\citet{Dutton2005,Gnedin2007} and \citet{Dutton2008}, potential
solutions to this problem include: (i) revising the adiabatic
hypothesis so as to reduce (or even reverse!) the halo contraction;
(ii) adopting lighter stellar mass-to-light ratios in order to
minimize the contribution of baryons and to allow for more dark mass
enclosed within disk galaxies; or (iii) modifying the cosmological
parameters so as to reduce halo concentrations. We shall concentrate
our analysis here on option (i), although we note that all three
possibilities are probably equally important and should be treated on
equal footing when addressing these issues in semianalytic models of
galaxy formation.

The possibility that the actual response of the halo to the assembly of
the disk might differ substantially from the predictions of the
adiabatic contraction formalism has been noted
before. \citet{Barnes1987} and \citet{Sellwood1999} have remarked that
the adiabatic contraction hypothesis might lead to substantial
overestimation of the halo compression. As discussed by
\citet{Sellwood2005} and \citet{Choi2006}, such deviations are likely
to depend strongly on the orbital structure of a halo, performing best
when most particles have large tangential motions, but poorly in
systems with radially anisotropic velocity distributions.

It should also be pointed out that the studies mentioned above were
carried out by perturbing simple spherical models with a potential
term designed to imitate the growth of an assembling galaxy. These
studies, therefore, miss the hierarchical nature of the assembly of a
galaxy and of its surrounding halo. The study of \citet{Gnedin2004},
on the other hand, uses the proper cosmological context, but focusses
on simulations of galaxy clusters, rather than the galaxy-sized
systems of interest for the issues discussed above. In this respect,
our study is similar to that of \citet{Gustafsson2006}, who studied
four simulations of galaxy-sized halos in the $\Lambda$CDM scenario.

The assembly of a central galaxy also modifies the three-dimensional
shape of the surrounding dark matter halo. This was already noted in
the first simulations to include, in addition to a dark matter halo,
a dissipative baryonic component, such as the early work of
\citet{Katz1991} and \citet{Katz1993}. \citet{Dubinski1994} studied
this further by growing adiabatically a central mass concentration
inside a triaxial dark matter halo and confirmed that the steepening
of the potential leads to much rounder halo shapes. This result was
confirmed by Kazantzidis et al (2004), who also noticed the effect in
their cluster simulations. We extend this body of work by focussing,
as in \citet{Hayashi2007}, on the shape of the gravitational potential
rather than on the axial ratios of the inertia tensor, as well as on
its dependence on radius.

The plan of this paper is as follows: \S~\ref{sec:numexp} describes
the numerical simulations and \S~\ref{sec:res} presents the main
results. After a brief general description of the evolution in
\S~\ref{sec:genev}, \S~\ref{sec:mj} describes the main properties of
the simulated central galaxies; \S~\ref{sec:shape} discusses the
modifications they induce on the shape of the halo gravitational
potential; while \S~\ref{sec:adcont} compares the mass profiles of
dark halos before and after the inclusion of a dissipative baryonic
component. We apply these results to the Milky Way in \S~\ref{sec:MW}
and compare with earlier work in \S~\ref{sec:comp}. We end with a
brief summary of our main conclusions in \S~\ref{sec:conc}.

\begin{figure}
\begin{center}
\includegraphics[width=0.9\linewidth,clip]{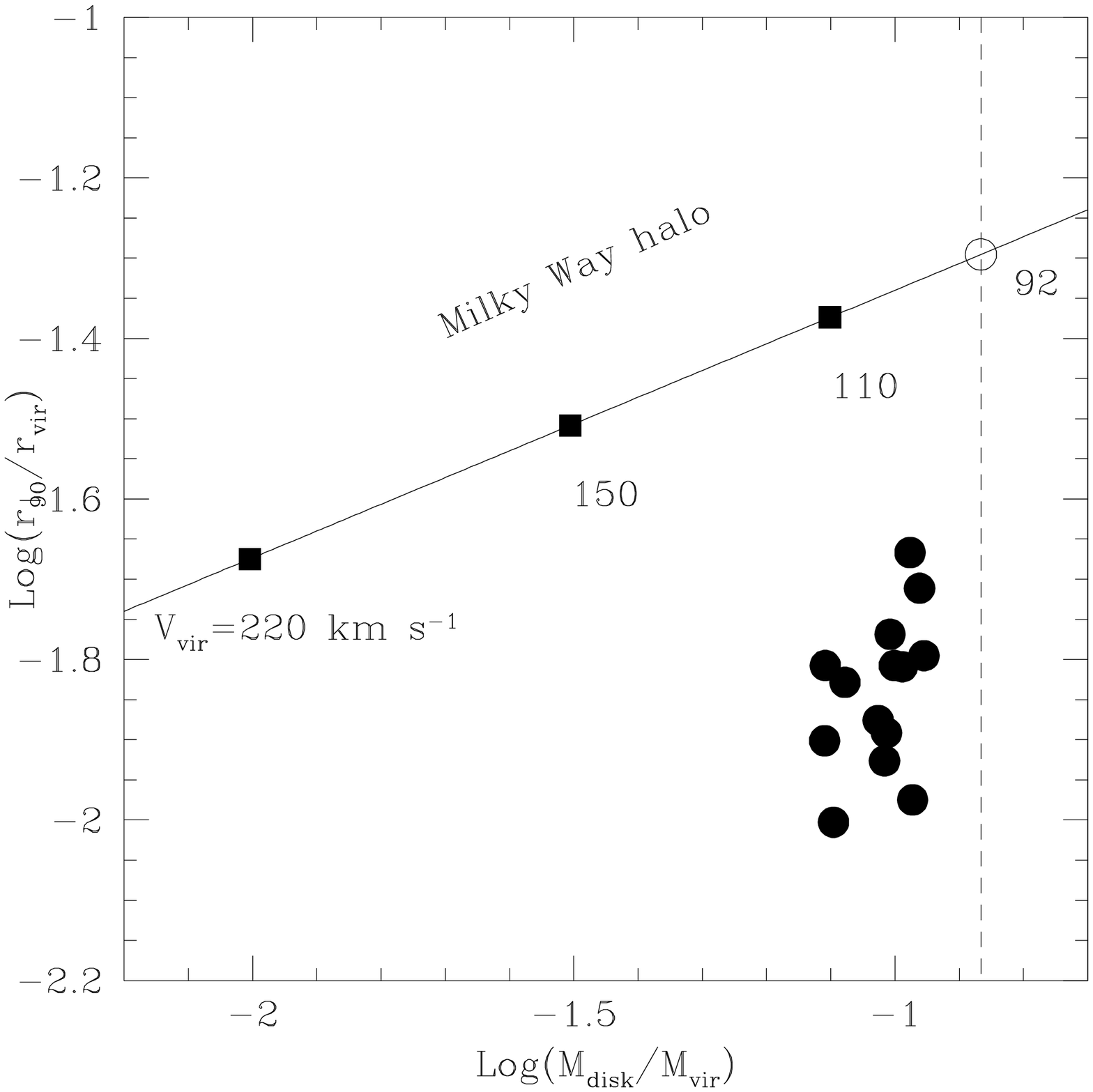}
\end{center}
\caption{Radius containing 90\% of the baryonic mass, $r_{90}$, versus
  the total baryonic mass of the central galaxy, $M_{\rm disk}$, in
  our simulated halos (filled circles), expressed in units of the halo
  virial values. The dashed vertical line the universal baryon
  fraction, $f_{\rm bar}=\Omega_b/\Omega_{\rm M}$. About $70$-$80\%$
  of all baryons have been collected in the central galaxy. The solid
  curve shows the corresponding values for the Milky Way, varying the
  virial mass of its halo. Symbols along the line indicate a few
  different values of the virial velocity, $V_{\rm vir}=220$,
  $150$, $110$, and $91$ km/s.  Note that the simulated central
  galaxies are substantially more massive and smaller than spirals
  like the Milky Way. This maximizes the effect of baryons in
  transforming the shape and mass profile of the dark halo.
\label{figs:r90}}
\end{figure} 
\begin{figure}
\begin{center}
\includegraphics[width=0.9\linewidth,clip]{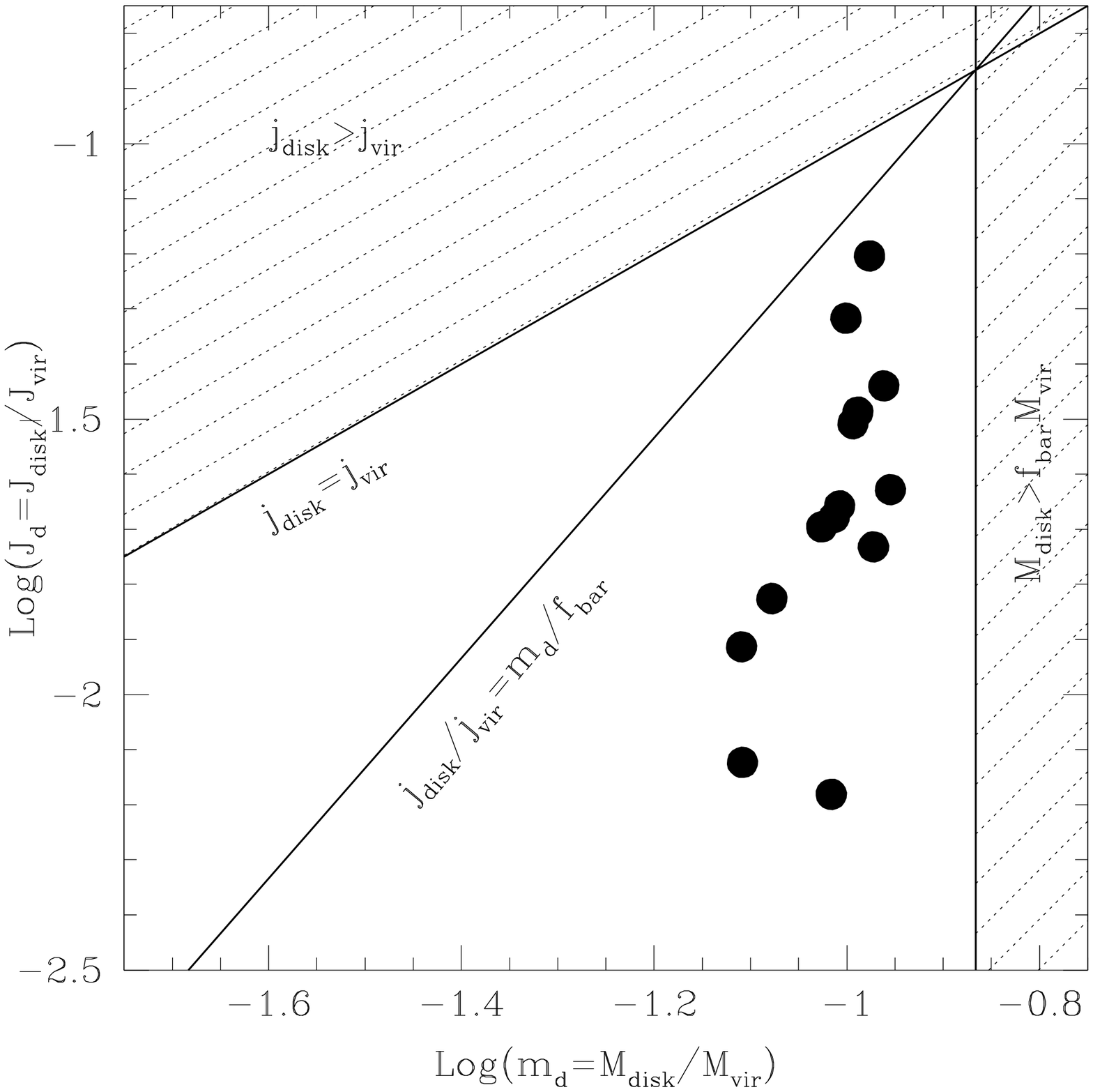}
\end{center}
\caption{Angular momentum and mass of the central galaxy in our
  simulations, expressed in units of the virial values. The curve
  labeled $j_{\rm disk}=j_{\rm vir}$ corresponds to central galaxies
  with the same {\it specific} angular momentum as the surrounding
  halo (a common assumption of semianalytic models). Those on the
  curve labeled $j_{\rm disk}/j_{\rm vir}=m_d/f_{\rm bar}$ have
  specific angular momenta (in units of the halo's) that scale in
  proportion to the fraction of baryons within the virial radius that
  have collected in the central galaxy. Note that the simulated
  central galaxies have even lower angular momenta than in the latter
  assumption, highlighting the large angular momentum losses that
  accompany the assembly of the central galaxy, and explaining the
  small sizes of the disks shown in Fig.~\ref{figs:r90}.
\label{figs:mj}}
\end{figure} 

\section{Numerical Simulations}
\label{sec:numexp}

\subsection{Cosmological parameters}

We adopt cosmological parameters consistent with the combined analysis
of the 2dfGRS \citep{Colless2001} and the first-year WMAP data
\citep{Spergel2003}: a present day value of the Hubble constant of
$H_0=70$ km/s/Mpc; a scale-free initial density perturbation spectrum
with no tilt and normalized by the linear rms mass fluctuations on $8
h^{-1}$ Mpc spheres, $\sigma_8=0.9$. The matter-energy content of the
Universe is expressed in units of the present-day critical density for
closure, and contains a dominant cosmological constant term,
$\Omega_{\Lambda}=0.7$, as well as contributions to the matter
content, $\Omega_{\rm M}=0.3$, from cold dark matter (CDM),
$\Omega_{\rm CDM}=0.259$, and baryons, $\Omega_{\rm b}=0.041$.

\subsection{Code and initial conditions}

We have performed a suite of 13 numerical simulations of the formation
of galaxy-sized halos in the $\Lambda$CDM cosmogony. Each simulation
follows the evolution of a relatively small region of the Universe,
excised from a $432^3$-particle simulation of a large $50 \, h^{-1}$
Mpc periodic box \citep{Reed2003}, and resimulated at much higher mass
and spatial resolution. Each ``zoomed-in'' re-simulation follows the
formation of a single galaxy-sized halo of mass $\sim 10^{12}\,
M_{\odot}$ and its immediate surroundings. The simulations include the
tidal fields of the parent simulation, and follows the coupled
evolution of gas and dark matter. The hydrodynamical evolution of the
gaseous component is followed using the Smooth Particle Hydrodynamics
(SPH) technique. This re-simulation technique follows standard
practice, as described, for example, by \citet{Power2003}. All our
simulations were performed using GASOLINE, a parallel N-body/SPH code
described in detail by \citet{Wadsley2004}.

\subsection{Halo selection and resimulation}

The $13$ halos were selected from a list of all $\sim 10^{12}\,
M_{\odot}$ halos in the parent simulation, with a mild bias to avoid
objects that have undergone a major merger after $z=1$ or that have,
at $z=0$, unusually low ($\lambda<0.03$) spin parameter.  The mass
resolution of the resimulations is such that halos are represented
with $\sim 40,000$ dark matter particles within their virial radius
{\footnote{ We define the {\it virial} radius, $r_{\rm vir}$, of a
system as the radius of a sphere of mean density $\Delta_{\rm vir}(z)$
times the critical density for closure. This definition defines
implicitly the virial mass, $M_{\rm vir}$, as that enclosed within
$r_{\rm vir}$, and the virial velocity, $V_{\rm vir}$, as the circular
velocity measured at $r_{\rm vir}$. Quantities characterizing a system
will be measured within $r_{\rm vir}$, unless otherwise specified. The
virial density contrast, $\Delta_{\rm vir}(z)$ is given by
$\Delta_{\rm vir}(z)=18\pi^2+82f(z)-39f(z)^2$, where
$f(z)=[\Omega_0(1+z)^3/(\Omega_0(1+z)^3+\Omega_\Lambda))]-1$ and
$\Omega_0=\Omega_{\rm
CDM}+\Omega_{b}$\citep{bryanandnorman98}. $\Delta_{\rm vir}\approx 100$
at $z=0$.}}
at $z=0$. Two sets of simulations are performed for each halo, one
where only a dark matter component is followed, and another where dark
matter and baryons are included. Dark matter-only simulations (``DMO''
for short) assume that the total matter content of the Universe is in
cold dark matter, $\Omega_{\rm M}=\Omega_{\rm CDM}=0.3$. 

Simulations including a baryonic component (``DM+B'' for short)
include equal number of gas and dark matter particles, with masses
modified to satisfy our choice of $\Omega_b=0.041$ and $\Omega_{\rm
  CDM}=0.259$. The gaseous component is evolved using SPH including,
besides the self-gravity of gas and dark matter, pressure gradients
and shocks.  Baryons are allowed to cool radiatively according to the
cooling function of a gas of primordial composition down to a
temperature of $10^4$ K, below which cooling is disabled. No star
formation or feedback is included. As we discuss below, this choice
maximizes cooling and favors the collection of most baryons at the
center of each dark halo. Although unrealistic as a model for galaxy
formation, this choice has the virtue of simplicity and also allows us
to examine the halo response when the effect of the baryonic component
is maximal.

Pairwise gravitational interactions are softened adopting a spline
softening length kept fixed in comoving coordinates.  We test for
numerical resolution effects by simulating each halo with 8 times
fewer particles. In the interest of brevity we do not present results
from this low-resolution series here, but we have checked explicitly
that none of the conclusions we present here are modified by this
change in numerical resolution. We have also increased the number of
particles in one case. Halo S02h is equivalent to S02, but with $\sim
3.4$ times more particles: at $z=0$ this halo has $\sim 140,000$
particles per component within the virial radius.

Table~\ref{tab:sims} lists the main properties of the simulated halos.

\begin{figure*}
\begin{center}
\includegraphics[width=1.0\linewidth,clip]{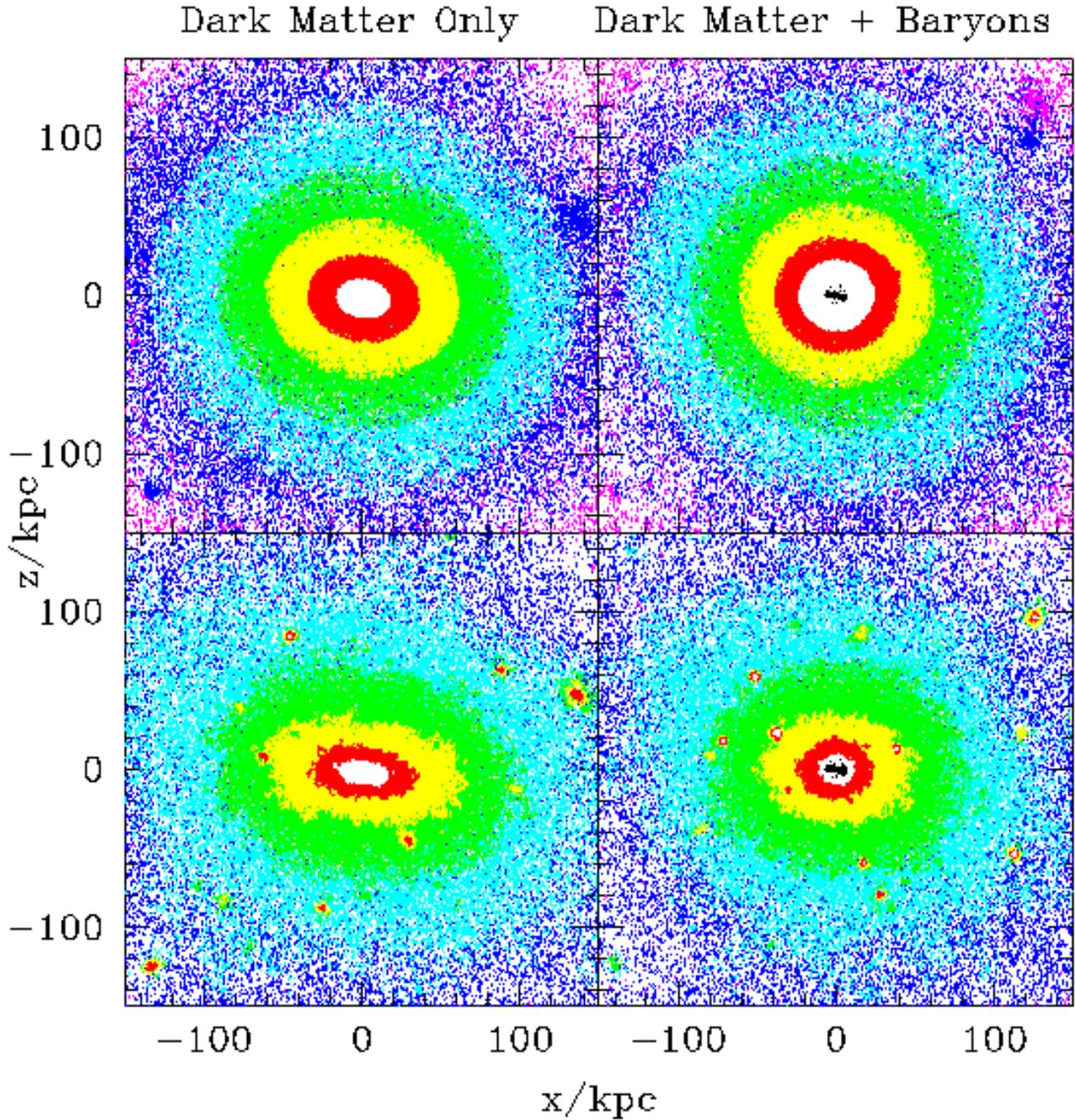}
\end{center}
\caption{Dark matter particles corresponding, at $z=0$, to the two
  resimulations of halo S02h, ``dark matter only'' (left panels) and
  ``dark matter plus baryons'' (right panels). Particles are colored
  according to the value of the gravitational potential (binned in
  logarithmic units, top panels) or the local density (bottom
  panels). Note that the halo responds to the assembly of the central
  galaxy (shown in black in the right-hand panels) by becoming
  noticeably more spherical. Projections are chosen so that the
  rotation axis of the central disk coincides with the z-axis. The
  central disk is shown with black dots, to illustrate the orientation
  of the disk relative to the halo shape. In general, the central disk
  is well aligned with the minor axis of the halo.
\label{figs:xyisopot}}
\end{figure*}

\begin{figure}
\begin{center}
\includegraphics[width=\linewidth,clip]{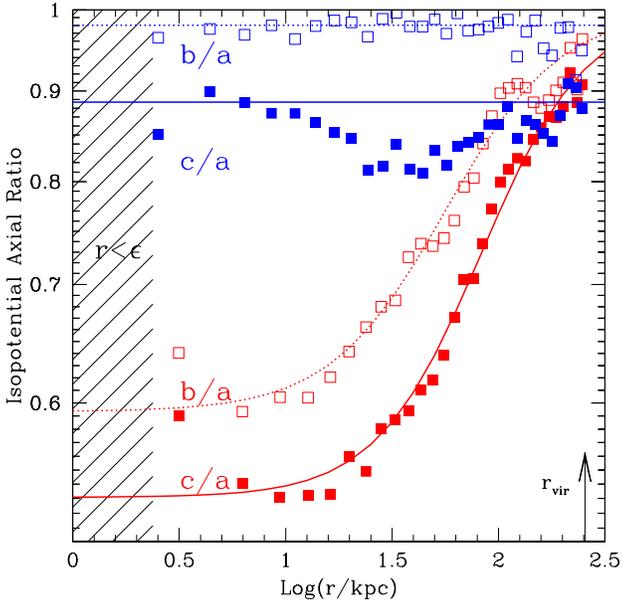}
\caption{"Radial" 
{\bf $(r=\sqrt{a^2+b^2+c^2})$}
dependence of axial ratios $b/a$ (open symbols) and
  $c/a$ (filled symbols) for our high-resolution run S02h. Halo shapes
  are measured by fitting 3D ellipsoids to the position of dark matter
  particles in narrow logarithmic bins of the gravitational potential.
  The potential is computed using {\it only} dark matter particles, so
  it actually corresponds to the contribution of the dark matter
  component to the overall potential. Red (lower) symbols correspond
  to the DMO run; blue (upper) symbols to the DM+B run. Note that the
  central galaxy turns a rather triaxial, nearly prolate halo into an
  axisymmetric, nearly oblate one. Curves are fits using
  eq.~\ref{eq:tanhfit}. 
  Fit parameters for this halo and the mean value computed over the sample are
  listed in Table~\ref{tab:fitpar}.
\label{figs:axrats02h}}
\end{center}
\end{figure} 
\begin{figure}
\begin{center}
\includegraphics[width=\linewidth,clip]{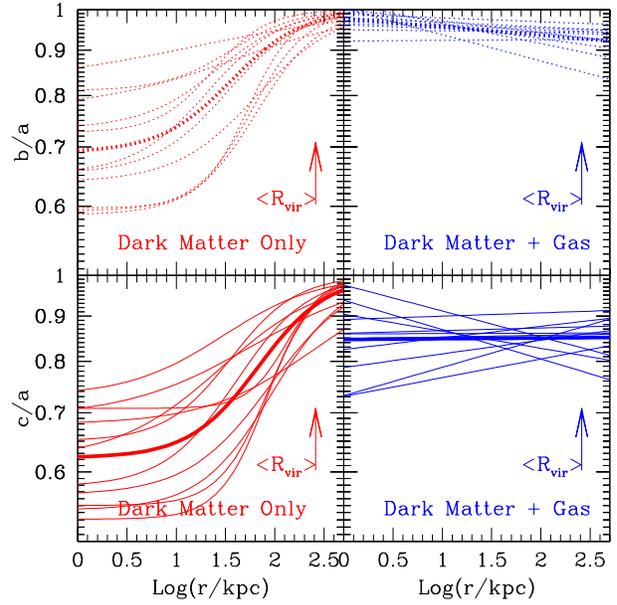}
\caption{"Radial"
{\bf
$(r=\sqrt{a^2+b^2+c^2})$ 
}
dependence of axial ratios $b/a$ (upper panel) and
  $c/a$ (lower panel) for all simulated haloes.  The curves for the
  ``dark matter only'' runs are computed by fitting
  eq.~\ref{eq:tanhfit} to the radial dependence of the axial ratios of
  particles along isopotential contours. The contours in all cases
  refer to the potential contributed solely by the dark matter
  component. Note that, as a result of the assembly of the central
  galaxy, (i) halos become nearly oblate and (ii) axial ratios become
  approximately independent of radius. The effect on the shape of the
  halo extends almost to the virial radius, far beyond the actual size
  of the central galaxy. Thick lines correspond to mean values computed
  over the sample.
\label{figs:fitall}}
\end{center}
\end{figure} 

\section{Results}
\label{sec:res}

\subsection{General evolution}
\label{sec:genev}

Because our simulations neglect the formation of stars and their
feedback, the evolution of the baryonic component is characterized by
the rapid cooling and collapse of baryons at the center of the early
collapsing progenitors of the final halo. As a result, the main mode
of galaxy assembly is mergers: up to $70\%$ of all baryons in the
central galaxy were accreted in the form of dense, cold, gaseous
clumps that sink to the center through dynamical friction.

As shown in earlier work, \citep[see,
  e.g.,][]{Navarro1991,Navarro1994,Navarro1997}, this mode of assembly
leads to very efficient accretion of baryons into the central galaxy
and to the transfer of much of their angular momentum to the
surrounding halo. Because baryons are assumed to remain in gaseous
form at all times, those that can cool are forced to settle into
centrifugally-supported disks at the center of their surrounding dark
halos. Thus, we expect the central galaxies in our simulations to be
disks that contain a large fraction of all baryons within the virial
radius, and to have angular momenta well below the angular momentum
content of the system as a whole.

Figure~\ref{figs:xyall} illustrates this for the case of our highest
resolution halo, S02h (see Table~\ref{tab:sims}). The four panels in
this figure show the dark matter (black) and baryonic (colored)
components, zooming by consecutive factors of 3 toward the central
galaxy. Cold gas ($T<10^{4.5}$K) is shown in green, whereas hot gas
($T>10^{4.5}$K) is shown in magenta. Note that most of the cold gas
inhabits the center of the main halo and its substructures, where it
forms easily identifiable thin, massive disks. All cold baryons
associated with the central disk are contained within a sphere of
radius $r_{\rm glx}=10$ kpc, which we shall use hereafter to define
the central galaxy in all runs.

\subsection{Mass, size and angular momentum of central galaxies}
\label{sec:mj}

The central galaxy contains most of the baryons within the virial
radius of the system ($74\%$ in the case of S02h, see
Table~\ref{tab:sims}). This is shown in Fig.~\ref{figs:r90}, where we
plot, for all our simulations, the baryonic mass of the central
galaxy, $M_{\rm disk}$, versus the radius, $r_{90}$, that contains
90\% of its mass, scaled to the virial mass and radius of the system,
respectively. The vertical dashed line indicates the universal baryon
fraction in the simulations, $f_{\rm bar}=\Omega_{\rm bar}/\Omega_{\rm
  M}=0.137$.  Typically, $70$ to $80\%$ of all baryons within $r_{\rm
  vir}$ are found in fairly small central disks that are fully
contained within a radius of order $\sim 3\%$ of the virial radius.

We may compare this with a typical spiral galaxy like the Milky Way
(MW), where the mass and size of the baryonic component can be
estimated accurately. Assuming that the disk is exponential with total
mass $4.5 \times 10^{10}\, M_{\odot}$ and radial scalelength $R_d^{\rm
  MW}=2.5$ kpc and that the mass of the bulge is $M_{\rm bulge}=4.5
\times 10^{9}\, M_{\odot}$, we estimate that 90\% of the Milky Way
baryons are confined within $r_{90}^{\rm MW}=9.4$ kpc.

In order to compare this with the simulation results shown in
Fig.~\ref{figs:r90} we need make an assumption about the virial mass
of the Milky Way halo.  For a virial velocity of the order of the
rotation speed at the solar circle ($220$ km/s), the baryonic
component makes up only $\sim 1\%$ of the virial mass, or roughly
$7\%$ of all baryons within $r_{\rm vir}$. The simulated disks are
thus much smaller than a typical spiral like the MW. They are also
comparatively more massive.

Only if the virial velocity is as low as $110$ km/s do the MW
bulge+disk make up a fraction of available baryons as high as in our
simulations ($\sim 75\%$).  However, in that case $r_{90}^{\rm MW}\sim
0.06 r_{\rm vir}$, a factor of $\sim 4$ times larger than our
simulated disks.

The reason why simulated gaseous disks are so much smaller than
typical spirals is that baryons have lost a large fraction of their
angular momentum to the surrounding halo. This is illustrated in
Fig~\ref{figs:mj}. where we plot the angular momentum of the gaseous
disk (in units of that of the system as a whole, which is dominated by
the dark matter component), $J_d\equiv J_{\rm disk}/J_{\rm vir}=M_{\rm
  disk} \, j_{\rm disk}/M_{\rm vir} \, j_{\rm vir}$, versus the mass
fraction, $m_d\equiv M_{\rm disk}/M_{\rm vir}$.  These are the
parameters commonly adopted in semianalytic models of disk formation,
such as those of \citet{Mo1998}.

Since baryons acquire during the expansion phase as much angular
momentum as the dark matter, and it is unlikely that $M_{\rm
  disk}/M_{\rm vir}$ will exceed the universal baryon fraction, then
we do not expect the {\it specific} angular momentum of the disk,
$j_{\rm disk}$, to exceed that of the system as a whole, $j_{\rm
  vir}$. Therefore, central galaxies are unlikely to populate the
shaded areas of Fig.~\ref{figs:mj}.

Most semianalytic work assumes, for simplicity, that $j_{\rm
  disk}=j_{\rm vir}$ or, equivalently, that $J_d=m_d$, regardless of
the value of $m_d$. On the other hand, \citet{Navarro2000} argue that
it is unlikely that a galaxy where $m_d<<f_{\rm bar}$ may have the
same specific angular momentum as the whole system, and propose that
$j_{\rm disk}$, as a fraction of $j_{\rm vir}$, should be comparable
to the fraction of baryons that make up the central galaxy; i.e.,
$j_{\rm disk}/j_{\rm vir}=M_{\rm disk}/(f_{\rm bar} M_{\rm
  vir})=m_d/f_{\rm bar}$. This is illustrated in Fig.~\ref{figs:mj} by
the lower diagonal line.

Our simulated galaxies are well below both lines. This indicates that
baryons have specific angular momenta much lower than their
surrounding halos and explains their small sizes compared to typical
spirals. These are therefore unrealistic models of spiral galaxy
formation, but the combination of large mass and small size allows us
to probe the response of the dark halo in the case where the deepening
of the potential well due to the central galaxy is maximal.

\subsection{Halo shape}
\label{sec:shape}

As anticipated in \S~\ref{sec:intro}, the halo responds to the
presence of the central galaxy by becoming significantly more
spherical. This is illustrated in Fig.~\ref{figs:xyisopot}, where we
compare the shape of isodensity and isopotential contours for the
``dark matter only'' and ``dark matter plus baryons'' runs of halo
S02h. Panels on the left correspond to the DMO run, those on the right
to the DM+B run. Particles are colored according to their local values
of their gravitational potential or local density (binned in a
logarithmic scale), computed using {\it only} the dark matter
particles. Note that using the gravitational potential leads to much
more stable estimates of the shape of the halo, since it is much less
affected by the presence of substructures and other transient
fluctuations in the mass distribution. 

The isopotential contours are well approximated by ellipsoidal
surfaces, and we use the axial ratios of such ellipsoids to measure,
as a function or radius, the change in shape of halo S02h. We show the
result in Fig.~\ref{figs:axrats02h}. As discussed by
\citet{Hayashi2007}, the radial dependence of the axial ratios may be
approximated by the formula,
\begin{equation}
\log ({b\over a} \ {\rm or}\ {c\over a}) = \alpha \left[ \tanh \left(
  \gamma \log \frac{r}{r_{\alpha}} \right) -1 \right],
\label{eq:tanhfit}
\end{equation}
Here $\alpha$ parameterizes the central value of the axial ratio,
$(b/a)_0$ or $(c/a)_0$, by $10^{-2~\alpha}$; $r_{\alpha}$ indicates
the characteristic radius at which the axial ratio increases
significantly from its central value; and $\gamma$ regulates the
sharpness of the transition.

The presence of the central galaxy turns the halo from a triaxial,
nearly prolate system into an axisymmetric, nearly oblate system where
the axial ratio is nearly constant with radius. As a result,
eq.~\ref{eq:tanhfit} is not adequate for the nearly featureless radial
dependence of the axial ratios in the DM+B runs. We fit the latter
with a simple power-law,
\begin{equation}
\log ({b\over a} \ {\rm or}\ {c\over a}) =  \alpha \log r + \beta,
\label{eq:pwlawfit}
\end{equation}

The best-fit parameter to the mass and isopotential contour profile
shapes are given in Table~\ref{tab:fitpar}.

The profile fits for all simulations are compiled in
Fig.~\ref{figs:fitall}, and illustrate a few important points.  As a
result of the assembly of the central galaxy: (i) halos become nearly
oblate; (ii) axial ratios are roughly independent of radius; and (iii)
halo shapes are affected well beyond the size of the central galaxy,
and nearly as far out as the virial radius.

\subsection{Mass profile and contraction}
\label{sec:adcont}

\begin{figure}
\begin{center}
\includegraphics[width=\linewidth,clip]{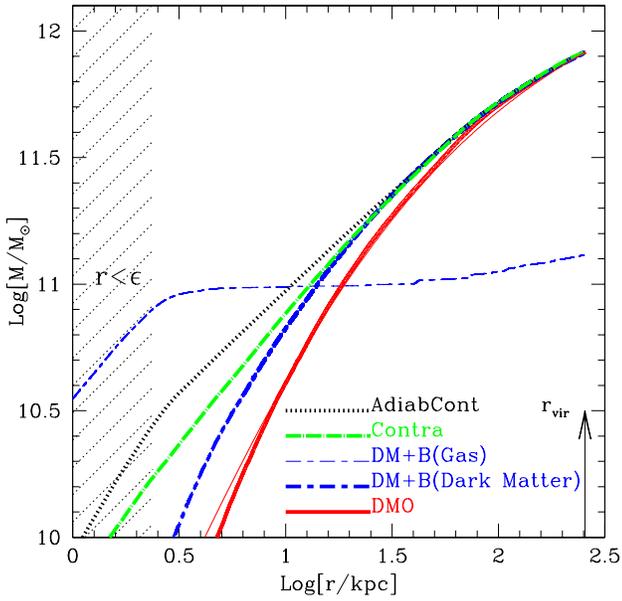}
\end{center}
\caption{ Enclosed mass profile of various components of halo
  S02h. The ``dark matter only'' profile is shown with a thick curve,
  after scaling masses by ($1-f_{\rm bar}$), so that, within $r_{\rm
    vir}$, the total dark mass of the DMO and DM+B runs will be
  comparable.  The thin line shows a Navarro-Frenk-White halo fit to
  the DMO profile. Other colors and line types correspond to the
  various components of the DM+B run, as specified in the figure
  labels. The thick dashed blue curve shows the dark mass profile for
  the DM+B run. The assembly of most baryons into a central galaxy
  (dotted magenta curve) has clearly led to a contraction of the dark
  mass profile. The profile predicted by the ``adiabatic contraction''
  formula (eq.~\ref{eq:adcont}; dot-dashed green curve) overpredicts
  the response of the halo. The modified adiabatic contraction model
  of \citet{Gnedin2004}, shown by the dotted cyan curve, also
  overestimates the halo response.
\label{figs:mprof}}
\end{figure}

Fig.~\ref{figs:mprof} shows the enclosed mass profile of the DMO and
DM+B runs corresponding to halo S02h. DMO dark masses have been scaled
by ($1-f_{\rm bar}$) so that the total dark mass in both the DMO and
DM+B runs are the same.  The DMO mass profile is shown by the thick
red solid curve and, as expected, it is well approximated by the NFW
\citep{Navarro1996,Navarro1997} formula (thin solid line). The central
disk that assembles in the DM+B run leads to a contraction of the dark
halo: there is {\it more} dark mass in the DM+B run {\it at all radii}
compared with the DMO run. This is a result common to all our
simulations; in no case do we see the halo ``expand'' as a results of
the assembly of the central galaxy.

The contraction, however, is not as pronounced as what would be
expected from the ``adiabatic contraction'' model discussed in
\S~\ref{sec:intro}. The adiabatic contraction prediction
(eq.~\ref{eq:adcont}) is shown by the thick dotted line in
Fig.~\ref{figs:mprof}. The discrepancy between model and numerical
results is not small. At $\sim 5$ kpc, a radius roughly twice the
gravitational softening and that encloses more than $1300$ dark
particles, the adiabatic-contraction formula predicts $\sim 2.5$ {\it
  times more dark mass than found in our simulations}. Even at a
radius of $10$ kpc, eq.~\ref{eq:adcont} overestimates the halo
contraction by more than $\sim 50\%$ in mass. The modified contraction
proposed by \citet{Gnedin2004} (see curve labeled ``Contra'' in
Fig.~\ref{figs:mprof}) fares better, but it still overpredicts the
results of the simulations.

\begin{figure}
\begin{center}
\includegraphics[width=\linewidth,clip]{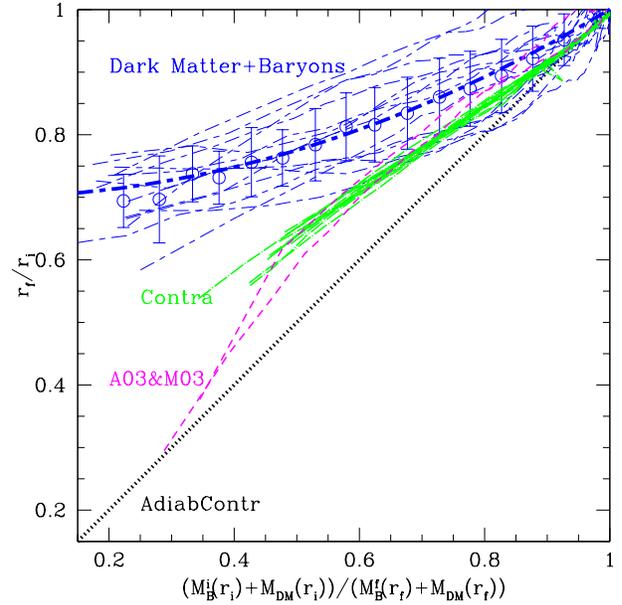}
\end{center}
\caption{Dark halo response to the assembly of a central galaxy. The
  ordinate shows the ratio, $r_f/r_i$, between the radius containing a
  given amount of dark mass in the DMO and DM+B runs, respectively,
  after scaling DMO masses by $1-f_{\rm bar}$. The smaller $r_f/r_i$
  the stronger the halo contraction. The x-axis shows the ratio
  between the total mass, $M_i$, contained within $r_i$ (in the DMO
  run) and $M_f$, that enclosed within $r_f$ (in the DM+B run). The
  adiabatic contraction formula (eq.~\ref{eq:adcont}) predicts that
  $r_f/r_i=M_i/M_f$; this is shown by the 1:1 line in the
  figure. Numerical results are shown for individual halos; from the
  radius that contains $1000$ dark particles outwards in order to
  minimize numerical uncertainties. The adiabatic-contraction formula
  overestimates the halo response. \citet{Gnedin2004}'s modified
  adiabatic contraction (``Contra'') does better but still
  overpredicts the halo response at most radii, and especially near
  the center. Symbols with error bars trace the median and quartiles
  of the numerical results. The thick upper curve is a fit using
  eq.~\ref{eq:fitcont}.
\label{figs:rm}}
\end{figure}

A similar result applies to all of our simulated halos. We
illustrate this in Fig.~\ref{figs:rm}, where we plot the ratio between
radii that contain a given number of dark matter particles in the DMO
run ($r_i$) and the DM+B run ($r_f$) versus the ratio between $M_i$,
the total mass within $r_i$ in the DMO run, and $M_f$, the total mass
within $r_f$ in the DM+B run. With this choice, the
adiabatic-contraction prediction is simply $r_f/r_i=M_i/M_f$, which is
traced by the 1:1 line in Fig.~\ref{figs:rm}.

The numerical simulations are shown by the upper curves in this
figure. Near the center, the baryons dominate over the DMO mass
profile, and therefore $M_i/M_f \ll 1$. Further out, the contribution
of baryons to the total enclosed mass decreases, approaching the
universal baryon fraction at the virial radius, where $M_i/M_f$ tends
to unity. The innermost radius plotted in each case corresponds to
that enclosing $1000$ dark matter particles. This is in all cases
comfortably larger than the gravitational softening, minimizing the
possibility that our results are unduly influenced by numerical
artifact. The open circles with error bars in Fig.~\ref{figs:rm} trace
the median and rms scatter of all our simulation results.

The ratio $r_f/r_i$ tends to a constant for $M_i/M_f \ll 1$,
suggesting that the halo response approaches saturation in regions
where baryons dominate. A simple formula captures the average
behaviour well (thick upper solid line),
\begin{equation}
{r_f / r_i} = 1 + a [(M_i/M_f)^n-1],
\label{eq:fitcont}
\end{equation}
with $a=0.3$ and $n=2$. As an application, we shall use this
expression below to explore what constraints this implies for the mass
and concentration of the halo of the Milky Way.

\subsection{Application to the Milky Way}
\label{sec:MW}

The Milky Way offers a case study for the results described
above. Because the mass and radial distribution of the baryonic
component, as well as the rotation speed of the local standard of rest
(LSR), are relatively well known, the total dark mass contained within
the solar circle is firmly constrained.
Adopting the same quantities for the bulge and disk of the Milky Way
adopted in \S~\ref{sec:mj}, we find that baryons contribute (in
quadrature) $\sim 171$ km/s to the circular velocity of the LSR, which we
assume to be $220$ km/s. This implies that the dark mass of the Milky
Way within the solar circle ($R_{\odot}=8$ kpc) is $\sim 3.5 \times
10^{10} \, M_{\odot}$. In order to allow for the possibility that some
of this dark mass may be baryonic, we shall treat this a formal upper
limit in the analysis that follows.

The second constraint comes from the total baryonic mass of the Milky
Way, under the plausible assumption that the mass of the central
galaxy cannot exceed the total mass of baryons within the virial
radius, $\approx f_{\rm bar} M_{\rm vir}$. Thus, the {\it minimum}
virial mass allowed for the MW halo by this constraint is $3.64 \times
10^{11}\, M_{\odot}$, which corresponds to a virial velocity of $\sim
92$ km/s.

\begin{figure}
\begin{center}
\includegraphics[width=\linewidth,clip]{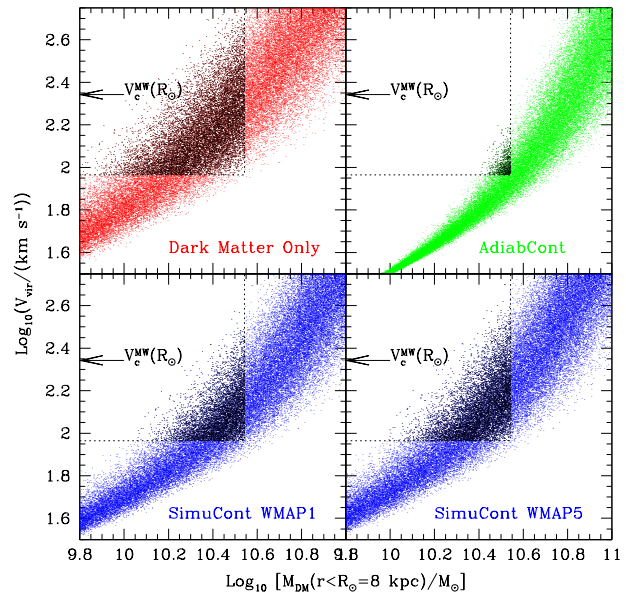}
\end{center}
\caption{Dark mass within $8$ kpc vs virial velocity for halos drawn
  at random from the $\Lambda$CDM mass-concentration relation
  (including scatter) of \citet{Neto2007}. The DMO panel (top left)
  shows the results neglecting the effects of baryons and assuming an
  NFW profile for the dark halos. The black ``wedge'' highlights the
  region of parameter space compatible with observations of the Milky
  Way. The panel labeled ``AdiabCont'' shows the result of contracting
  each halo adopting the adiabatic contraction formula
  (eq.~\ref{eq:adcont}). Very few halos would be consistent with the
  Milky Way if this formula holds. The bottom-left panels shows the
  result of applying our simulation results for the halo contraction,
  as captured by eq.~\ref{eq:fitcont}. The bottom-right panel repeats
  the same exercise, but after reducing all concentrations by $20\%$
  in order to mimic the expected concentration change resulting from
  adopting the latest cosmological parameters from the 5-year analysis
  of WMAP satellite data \citep{Duffy2008}.
\label{figs:vm8}}
\end{figure}

Are these constraints compatible with $\Lambda$CDM halos? 
The top left panel of Fig.~\ref{figs:vm8} shows the dark mass expected
within 8 kpc vs halo virial velocity.  Each dot in this panel
correspond to an NFW halo with concentration drawn at random from the
mass-concentration relation (including scatter) derived by
\citet{Neto2007} from the Millennium Simulation
\citep{Springel2005a}. As expected, the mass within 8 kpc increases
with the virial velocity (mass) of the halo, modulated by the fairly
large scatter in concentration at given halo mass.

The black ``wedge'' in this panel indicates the region allowed by the
MW constraints discussed above. As discussed by \citet{Eke2001}, most
$\Lambda$CDM halos satisfy the constraints, and would be consistent
with the MW if they were somehow able to avoid contraction. Note as
well that the larger the MW halo virial velocity the lower the allowed
concentration. For $V_{\rm vir}=220$ km/s (the value required by
semianalytic models to match the Tully-Fisher relation and the
luminosity function, see \S~\ref{sec:intro}) only halos with $c_{\rm
  vir}<9.5$ would be consistent with the Milky Way. For comparison,
the average concentration of $V_{\rm vir}=220$ km/s halos is $\langle
c_{\rm vir} \rangle \sim 10.4$ and its (lognormal) dispersion is
$\sigma_{\rm \log c}=2.9$, so this condition effectively excludes only
$57\%$ of such halos from the allowed pool.

The situation is rather different if halos are adiabatically
contracted (panel labeled ``AdiabCont'' in Fig.~\ref{figs:vm8}). The
contraction increases substantially the dark mass contained within 8
kpc, so that very few halos satisfy the MW constraints. For example,
essentially no halo with $V_{\rm vir}\approx 220$ km/s could host a
galaxy like the Milky Way, and the few that could would need to have
$c_{\rm vir} < 3.2$, many sigma away from the average concentration of
halos of that mass. The halo of the Milky Way would need to be a
very special halo of unusually low concentration if
adiabatic contraction holds.

Adopting the halo contraction of our numerical simulations improves
matters. This may be seen in the bottom-left panel of
Fig.~\ref{figs:vm8}, where we have contracted each halo using
eq.~\ref{eq:fitcont}. The range of allowed halo masses and
concentrations is broader; even some halos with $V_{\rm vir}\sim 220$
km/s could be consistent with the Milky Way, provided that $c_{\rm
  vir}<6.5$. This is significantly lower than the average at that
mass, but conditions are much less restrictive if the halo of the
Milky Way is less massive; for example, half of all $V_{\rm vir} \sim
130$ km/s halos satisfy the MW constraints. The possibility that the
virial velocity of the Milky Way is significantly lower than $220$ km/s has
indeed been advocated by a number of recent observational studies
\citep{Smith2007,Sales2007a,Xue2008}.

Lowering the average concentration at given mass also helps. This is
shown in the bottom-right panel of Fig.~\ref{figs:vm8}, where we have
repeated the exercise, but lowering all concentrations by
$20\%$. According to \citet{Duffy2008}, this is approximately the
change in average concentration when modifying the cosmological
parameters from the values we adopt here (which are consistent with
the first-year analysis of the WMAP satellite data) to those favoured
by the latest 5-year WMAP data analysis. With this revision, $18\%$ of
$V_{\rm vir}\sim 220$ km/s halos and roughly half of all $V_{\rm vir}\sim
160$ km/s halos would be consistent with the MW.

\subsection{Comparison with earlier work}
\label{sec:comp}

The failure of the adiabatic contraction formalism to match the
results of our simulations suggest that the response of the halo is
more complex than what can be captured with a simple model for the
contraction of spherical shells. This is confirmed by comparing our
results with earlier work.

Our results disagree not only with the traditional adiabatic
contraction formula, but also with the modified formalism proposed by
\citet{Gnedin2004} (shown with thin dash-dotted green lines in Fig.~\ref{figs:rm}).
These authors calibrated their results with numerical simulations not
dissimilar to ours, except for the mass scale---they used mainly
galaxy cluster halos. The mass scale, on the other hand, does not seem
to be the reason for the discrepancy. Our halo contraction is also
less pronounced than found in two galaxy-sized halo simulations, as
shown by the lines labeled A03\&M03 in Fig.~\ref{figs:rm}. These correspond
to gasdynamical simulations of galaxy formation presented by
\citet{Abadi2003a,Abadi2003b} and \citet{Meza2003}, and differ from
ours mainly in their inclusion of star formation and feedback
effects. Those two simulations seem to agree better with the
``Contra'' predictions than with our simulation results.  

Barring numerical artifact, these results seem to suggest that {\it
  the halo response does not depend solely on the initial and final
  distribution of baryons}. One possible explanation is that not only
{\it how much} baryonic mass has been deposited at the center of a
halo matters, but also the {\it mode} of its deposition. It is
certainly plausible that central galaxies assembled through merging of
dense subclumps may lead to different halo response than galaxies
assembled through a smooth flow of baryons to the center. Mergers of
baryonic subsystems may in principle pump energy into the dark halo
(through dynamical friction), altering its central structure and
softening its contraction. This possibility has been argued before
\citep[see, e.g.,][]{ElZant2001,Ma2004,Mo2004}, and seems to find
favor in our simulations, where mergers between massive baryonic
clumps are frequent and plentiful and halo contraction is less strong
than reported in earlier work.

Recent work has also speculated that mergers may lead to halo {\it
  expansion} and that this would help to reconcile the properties of
disk galaxies with $\Lambda$CDM halos \citep{Dutton2007}. Our halos
always contract, and it seems safe to conclude that our results
reflect the maximum effect of mergers on the halo response. We
conclude therefore that mergers alone are unlikely to result in halo
expansion.  If such expansion is truly needed to reconcile disk
properties with $\Lambda$CDM halos, it should come as a consequence of
other processes not considered here, such as feedback-driven winds
that may remove substantial fraction of baryons from the central
galaxy \citep[see, e.g.,][]{NavarroEkeFrenk1996,BabulFerguson1996}.

\section{Summary}
\label{sec:conc}

We have used a suite of cosmological N-body/gasdynamical simulations
to examine the modifications to the dark halo structure that result
from the assembly of a central galaxy. The formation of 13
$\Lambda$CDM halos is simulated twice, with and without a baryonic
(gaseous) component. For simplicity, the gasdynamic simulations
include radiative cooling but neglect star formation and
feedback. This favors the formation of massive central baryonic disks
at the center of the early collapsing progenitors of the final
halo. As these systems merge, the gaseous clumps merge and re-form a
disk at the center of the remnant.

The merger process leads to the transfer of a large fraction of the
angular momentum from the baryons to the halo. At $z=0$, the simulated
central galaxies are too massive and too small to be consistent with
observed spiral galaxies. Although unrealistic as a disk galaxy formation
model, these simulations allow us to probe the dark halo response in
the interesting case where the deepening of the potential well
resulting from the formation of the central galaxy is maximized. Our
main conclusions may be summarized as follows.

\begin{itemize}

\item
Dark halos become significantly more spherical as a result of the
assembly of the central galaxy. The triaxial, nearly prolate systems
that form in the absence of a baryonic component are transformed into
essentially oblate systems with a roughly constant axial isopotential
ratio $\langle c/a\rangle \approx 0.85$.

\item
Halos always contract in response to the formation of the central
galaxy. The ``adiabatic contraction'' formalism overestimates the halo
contraction in our simulations. The discrepancy increases toward the
centre, where the effect of baryons is larger. A simple
empirical formula (eq.~\ref{eq:fitcont}) describes our numerical
results. 

\item
The halo contraction in our simulations is also less pronounced than
found in earlier numerical work \citep[see, e.g.,][]{Gnedin2004}, and
suggest that the response of a halo does not depend on the final mass
and radial distribution of baryons in the central galaxy, but also on
the mode of their assembly.

\item
We apply these results to the Milky Way, where accurate estimates of
the mass of baryons and dark matter inside the solar circle
exist. These allow us to probe the range of halo virial mass and
concentration consistent with the constraints. Only halos of unusually
low mass and concentration would match such constraints if ``adiabatic
contraction'' holds.  

\item
This restriction is less severe if one uses the halo contraction
reported here (eq.~\ref{eq:fitcont}): although few $\Lambda$CDM halos
of virial velocity $\sim 220$ km/s would be eligible hosts for the
Milky Way, the situation improves for lower virial velocities. Halos
of average concentration with virial velocity as large as $V_{\rm
  vir}=135$ km/s, would be consistent with the MW constraints.

\end{itemize}

The dark halo response to the formation of a galaxy seems inextricably
linked to the full coupled evolution of baryons and dark matter and
may vary from system to system. Progress in our understanding of the
distribution of dark matter in a baryon-dominated system will thus
likely develop in step with our understanding of the particular
assembly history of each individual system.

\section*{Acknowledgments}

We thank James Wadsley, Joachim Stadel and Tom Quinn for allowing us
to use their excellent GASOLINE code for the numerical simulations
reported here.

\begin{table*}
\centering
\begin{tabular}{lrrrrrrrrr}
\hline
Label &$M_{\rm vir}$ & $r_{\rm vir}$ &$N_{\rm DMO}$ &$N_{DM}$ & $N_{\rm gas}$ &$N_{\rm gas}$&$j_{\rm DMO}$&$j_{\rm DM}$& $j_{\rm gas}$\\
      & [$10^{12}M_{\odot}$] & [kpc] &$(<r_{\rm vir})$ & $(<r_{\rm vir})$ & $(<r_{\rm vir})$ & $(<r_{\rm glx})$ & [kpc km s$^{-1}$] & [kpc km s$^{-1}$] & [kpc km s$^{-1}$]   \\\hline\hline
S01 & 1.22 & 277.82 &  53009 &  52684 &  54732 &  38260 & 1847.7 & 2162.4 &  462.0 \\
S02 & 0.94 & 254.18 &  40847 &  40687 &  39775 &  30613 & 1063.3 & 1088.3 &  332.8 \\
S02h& 0.95 & 255.89 & 139748 & 139729 & 140618 & 104309 & 1095.7 & 1200.3 &  359.1 \\
S03 & 1.51 & 298.24 &  65729 &  65376 &  66416 &  37445 & 2593.9 & 2883.4 &  438.1 \\
S04 & 0.85 & 246.02 &  37419 &  36432 &  39007 &  29534 & 1443.1 & 1626.7 &  508.8 \\
S05 & 0.99 & 258.64 &  42564 &  42334 &  45270 &  30266 & 2505.0 & 2543.2 &  173.3 \\
S06 & 1.22 & 278.05 &  53509 &  52652 &  55674 &  30411 & 2736.0 & 3056.0 &  289.4 \\
S07 & 0.86 & 246.95 &  37350 &  37305 &  36524 &  25743 & 1332.2 & 1385.1 &  310.0 \\
S08 & 0.84 & 245.39 &  35902 &  36224 &  38221 &  26031 & 1221.3 & 1292.5 &  277.6 \\
S09 & 0.84 & 245.23 &  36619 &  36259 &  37441 &  28264 &  850.5 &  929.5 &  531.8 \\
S10 & 0.97 & 257.68 &  42004 &  42074 &  43469 &  34475 & 1976.6 & 2231.8 &  430.7 \\
S11 & 0.88 & 249.53 &  40181 &  38190 &  39551 &  23579 & 1430.3 & 1676.8 &  289.6 \\
S12 & 0.90 & 251.25 &  39715 &  39341 &  38140 &  28738 &  715.7 &  764.0 &  381.5 \\
S13 & 1.08 & 266.72 &  46863 &  46337 &  50223 &  36708 & 1470.7 & 1607.5 &  261.3 \\

\hline\end{tabular}
\caption{Main parameters of simulated halos.  $M_{\rm vir}$ is the total
  (dark matter plus gas) mass inside the virial radius,
  $r_{\rm vir}$. $N_{\rm DMO}$ and $N_{\rm DM}$ is the number of dark
  matter particles inside the virial radius $r_{\rm vir}$ in the ``dark
  matter only'' and ``dark matter plus baryons'' simulations,
  respectively. Columns 6 and 7 list the number of gas particles,
  $N_{\rm gas}$, within the virial radius and within the radius used
  to define the central galaxy, $r_{\rm glx}=10$ kpc. The specific
  angular momentum of the dark matter within $r_{\rm vir}$ and of the
  baryons within $r_{\rm glx}$ is also listed. \label{tab:sims}}
\end{table*}

\begin{table*}
\centering
\begin{tabular}{lcccccccccc}
\hline
\multicolumn{1}{l}{} & \multicolumn{3}{c}{b/a (Dark Matter Only)} &  \multicolumn{3}{c}{c/a (Dak Matter Only)} & \multicolumn{2}{c}{b/a (Dark Matter+Gas)} & \multicolumn{2}{c}{c/a (Dark Matter+Gas)} \\
                  & $\alpha$ & $\gamma$ & $r_{\alpha}$  & $\alpha$ & $\gamma$ & $r_{\alpha}$ & $\alpha$ & $\beta$ & $\alpha$ & $\beta$ \\\hline\hline
S02h              & 0.113 & 1.767 &  58.31 &  0.137 &  2.102 &  83.13 & -0.00429 & -0.00508 &  0.00185 & -0.07127 \\
Sample Mean Values& 0.081 & 1.383 &  35.81 &  0.103 &  1.427 &  68.06 & -0.00934 & -0.01009 &  0.00081 & -0.07220 \\

\hline\end{tabular}
\caption{Best fitting parameters for isopotential contour profile shapes. 
Columns 1-3 (4-6)  correspond to axial ratio b/a (c/a) for Dark Matter Only runs (eq.~\ref{eq:tanhfit}). 
Columns 7-8 (9-10) correspond to axial ratio b/a (c/a) for Dark Matter+Gas  runs (eq.~\ref{eq:pwlawfit}). 
We quote values for the high resolution run S02h shown in figure~\ref{figs:axrats02h} and also the mean value for the sample 
shown as thick lines in figure~\ref{figs:fitall}.
   \label{tab:fitpar}}
\end{table*}

\bibliographystyle{mnras}
\bibliography{master}

\end{document}